\documentclass{article}
\usepackage{amsmath}

\input{tcilatex}

\begin{document}

\title{The Information Geometry of Space and Time\thanks{%
Presented at MaxEnt 2005, the 25th International Workshop on Bayesian
Inference and Maximum Entropy Methods (August 7-12, 2005, San Jose,
California, USA).}}
\author{Ariel Caticha \\
%EndAName
{\small Department of Physics, University at Albany--SUNY, Albany, NY 12222,
USA}}
\date{}
\maketitle

\begin{abstract}
Is the geometry of space a macroscopic manifestation of an underlying
microscopic statistical structure? Is geometrodynamics - the theory of
gravity - derivable from general principles for processing information?
Tentative answers are suggested by a model of geometrodynamics based on the
statistical concepts of entropy, the Fisher-Rao information metric, and
entropic dynamics. The model shows remarkable similarities with the 3+1
formulation of general relativity. For example, the dynamical degrees of
freedom are those that specify the conformal geometry of space; there is a
gauge symmetry under 3d diffeomorphisms; there is no reference to an
external time; and the theory is time reversible. There is, in adition, a
gauge symmetry under scale transformations. I conjecture that under a
suitable choice of gauge one can recover the usual notion of a relativistic
space-time.
\end{abstract}

\section{Statistical Geometrodynamics?}

The point of view that has been prevalent among scientists is that the laws
of physics mirror the laws of nature. The reflection might be imperfect, a
mere approximation to the real thing, but it is a reflection nonetheless.
The connection between physics and nature could, however, be less direct.
The laws of physics could be mere rules for processing information about
nature. If this second point of view turns out to be correct one would
expect many aspects of physics to mirror the structure of theories of
inference. Indeed, it should be possible to derive the \textquotedblleft
laws of physics\textquotedblright\ appropriate to a certain problem by
applying standard rules of inference to the information that happens to be
relevant to the problem at hand.\footnote{%
Basic requirements of consistency, objectivity, universality, and honesty
lead to the theory of probability \cite{Cox46}\cite{Jaynes03} and to the
method of maximum entropy \cite{Jaynes57}--\cite{Caticha03a} as the uniquely
natural rules of inference.}

There is strong evidence that this second point of view is worth pursuing.
For example, most of the formal structure of statistical mechanics can be
explained as a consequence of the method of maximum entropy \cite{Jaynes57}.
A second example is given by quantum mechanics. It is less well-known but
nevertheless still true that many features of the quantum formalism that are
usually introduced as postulates (the Hilbert spaces, linear and unitary
time evolution, the Born probability rule, Hermitian observables, etc.) can
be derived from principles of inference (consistency, entropy, and so on)
once the subject matter has been correctly identified \cite{Caticha98}.

This paper explores the possibility that the general theory of relativity is
also a theory of this type; that it can be derived from an underlying
\textquotedblleft statistical geometrodynamics\textquotedblright\ in much
the same way that thermodynamics can be explained from an underlying
statistical mechanics.

Our subject can be approached from a different direction. Modern
developments in statistical inference \cite{Amari00}\cite{Rodriguez02} have
shown that geometrical concepts turn out to be extremely natural tools to
manipulate information. If physics is nothing but manipulating information
about the world, then this suggests an explanation for the central role that
geometry has always played in physics. It also suggests that it should be
possible to explain basic geometrical notions such as spatial distance and
temporal duration in terms of even more basic statistical notions.

In section 2 we take the first step towards specifying the subject matter.
(The statistical geometrodynamics developed here is a model for empty
vacuum; it does not include matter.) The difficulty is that space and time
are invisible. What we see is not space but matter in space and it is not
clear how to disentangle which properties should be attributed to matter and
which to space. The best one can do is sprinkle space with ideal test
particles that are neutral to all interactions and are describable by a
minimal number of attributes. Such purest form of matter is a dust of
identical particles; they only interact gravitationally, and being identical
the only attribute that distinguishes them is their position.

Then we introduce the main assumption: there is an intrinsic fuzziness to
space which is revealed by an irreducible uncertainty in the location of the
test particles. Thus, to each point in space we associate a probability
distribution. The overall state of space -- the macrostate -- is defined by
the product of the distributions associated to the individual points. The
geometry of space is the geometry of all the distances between test
particles and this geometry is of statistical origin \cite{Caticha03b}.
Identical particles that are close together are easy to confuse, those that
are far apart are easy to distinguish. The distance between two neighboring
particles is the distinguishability distance between the corresponding
probability distributions which is given by the Fisher-Rao metric \cite%
{Fisher25}. A remarkable feature of this choice of distance is its
uniqueness: the Fisher-Rao metric is the only metric that takes account of
the fact that we deal with probability distributions and not with
\textquotedblleft structureless\textquotedblright\ points \cite{Cencov81}. A
second remarkable feature is that the information geometry we introduce does
not define the full Riemannian geometry of space but only its conformal
geometry. This appears at first to be a threat to the whole program but it
turns to be just what we need in a theory of gravity \cite{York71}.

But the task of specifying the subject matter is not yet finished. A proper
understanding of what we mean by a state requires that we be able to
quantify the extent to which one state can be distinguished from another. In
particular, the measure of time and dynamics itself derive from our capacity
to measure change between one state that we call `earlier' and another state
that we call `later' \cite{Caticha01}. In section 3 we measure the change
from one state to another using, once again, the Fisher-Rao metric. A
peculiarity that arises when comparing the states of systems with a
continuum of degrees of freedom turns out to be very significant. In such
cases we have to make an explicit choice about which location in the later
state corresponds, matches, or ultimately, is \emph{the same as} a given
location in the earlier state. The method of maximum entropy provides a
natural criterion to achieve the best match between two successive states%
\footnote{%
To borrow a term coined by Barbour, we might say that best-matching
establishes a relation of \textquotedblleft equilocality\textquotedblright\ 
\cite{Barbour94}.}. The resulting best-matching condition closely resembles
the diffeomorphism constraint in the Hamiltonian formulation of general
relativity \cite{Arnowitt62}.

It is interesting that the Fisher-Rao metric is used in two ways that are
conceptually very different. One is to distinguish neighboring points, the
other to distinguish successive states. The first is related to spatial
distance, the second to temporal duration. This suggests an explanation of
the old puzzle of how can space and time be so different physically and yet
be represented mathematically in such a symmetrical way.

Having specified the states, in section 4 we tackle the dynamics. We ask:
Given the initial and the final states, what trajectory is the system
expected to follow? In the usual approach the dynamics is postulated. No
further explanation is needed because \textquotedblleft that's the way
nature is.\textquotedblright\ But this route is not open to us. We are just
making inferences from relevant information and the expected trajectory is
obtained, without additional postulates, from a principle of inference: the
method of maximum entropy \cite{Caticha03b}\cite{Caticha02}.

The resulting entropic\ dynamics is not identical with the general theory of
relativity but there are remarkable similarities which strongly suggest that
general relativity can be obtained in some appropriate limit.

\section{The information geometry of space}

Consider a cloud of identical test particles -- specks of dust -- suspended
in an otherwise empty space. There are no rulers and no clocks, just dust.
Being identical the particles are easy to confuse. The only distinction
between two of them is that one happens to be here while the other is over
there. To distinguish one speck of dust from another we assign labels or
coordinates to each particle. We assume that three real numbers $%
(y^{1},y^{2},y^{3})$ are sufficient.

But particles can be mislabeled. Then the \textquotedblleft
true\textquotedblright\ coordinates $y$ are unknown and one can only provide
an estimate, $x$. Let $p(y|x)dy$ be the probability that the particle
labeled $x$ should have been labeled $y$. The labels $x$ are introduced to
distinguish one particle from another, but can we distinguish a particle at $%
x$ from another at $x+dx$? If $dx$ is small enough the corresponding
probability distributions $p(y|x)$ and $p(y|x+dx)$ overlap considerably and
it is easy to confuse them. We seek a quantitative measure of the extent to
which these two distributions can be distinguished.

The following crude argument is intuitively appealing. Consider the relative
difference, 
\begin{equation}
\frac{p(y|x+dx)-p(y|x)}{p(y|x)}=\frac{\partial \log p(y|x)}{\partial x^{{}i}}%
\,dx^{{}i},
\end{equation}%
where we adopt Einstein's convention of summing over the repeated indices.
The expected value of this relative difference does not provide us with the
desired measure of distinguishability because it vanishes identically.
However, the variance 
\begin{equation}
d\lambda ^{2}=\int d^{3}y\,p(y|x)\,\frac{\partial \log p(y|x)}{\partial
x^{{}i}}\,\frac{\partial \log p(y|x)}{\partial x^{{}j}}\,dx^{{}i}dx^{{}j}%
\overset{\limfunc{def}}{=}\gamma _{ij}\,(x)dx^{{}i}dx^{{}j}\,\,.
\label{Fisher metric}
\end{equation}%
is positive definite -- it vanishes if and only if $dx^{i}=0$. This is the
measure of distinguishability we seek. Except for an overall multiplicative
constant, the Fisher-Rao metric $\gamma _{ij}$ is the only Riemannian metric
that adequately reflects the underlying statistical nature of the manifold
of distributions $p(y|x)$ \cite{Cencov81}. An important property that will
be exploited below is the relation between the metric (\ref{Fisher metric})
and the entropy of $p(y|x+dx)$ relative to $p(y|x)$, 
\begin{equation}
S[p(y|x+dx)|p(y|x)]=-\int d^{3}y\,p(y|x+dx)\,\log \frac{p(y|x+dx)}{p(y|x)}=-%
\frac{1}{2}d\lambda ^{2}.  \label{Sdlambda}
\end{equation}%
Thus, maximizing the relative entropy $S$ is equivalent to minimizing the
distance $d\lambda ^{2}$.

We take the further step of interpreting $d\lambda $ as the \emph{spatial}
distance. Indeed, one would normally say that the reason it is easy to
confuse two particles is that they happen to be too close together. We argue
in the opposite direction and \emph{explain} that the reason the particles
at $x$ and at $x+dx$ are close together is \emph{because} they are difficult
to distinguish.

The origin of the uncertainty is left unspecified. We assume, however, that
any two particles at the same location in space are affected by the same
irreducible uncertainty. Then the uncertainty is not linked to the particle,
but to the place: the source of the uncertainty is a noise, a fluctuation or
a fuzziness in space itself.

To assign an explicit $p(y|x)$ we consider what is perhaps the simplest
possibility. We assume that $p(y|x)$ is sharply localized in a small
neighborhood about $x$ and that within this very small region curvature
effects can be neglected. We further assume that the information that is
relevant to our problem is given by the expected values $\langle
y^{i}\rangle =x^{i}$ and the covariance matrix $\langle
(y^{i}-x^{i})(y^{j}-x^{j})\rangle =C^{ij}(x)$. This is physically
reasonable: for each test particle we have estimates of its position and of
a small margin of error. Since the underlying space is locally flat, $p(y|x)$
can be determined maximizing entropy relative to a uniform measure. This
leads to a Gaussian distribution, 
\begin{equation}
p(y|x)=\frac{C^{1/2}}{(2\pi )^{3/2}}\,\exp \left[ -\frac{1}{2}%
C_{ij}(y^{i}-x^{i})(y^{j}-x^{j})\right] ,
\end{equation}%
where $C_{ij}$ is the inverse of the covariance matrix $C^{ij}$, $%
C^{ik}C_{kj}=\delta _{j}^{i}$, and $C=\det C_{ij}$. The corresponding metric
is obtained from eq.(\ref{Fisher metric}). For small uncertainties $C_{ij}(x)
$ is constant within the region where $p(y|x)$ is appreciable and we get $%
\gamma _{ij}(x)=C_{ij}(x)$. The metric changes smoothly over space and, in
general, space is curved. Connections, curvatures, and other aspects of the
geometry can be computed in the standard way.

To summarize, to each point $x$ in space we associate a probability
distribution,

\begin{equation}
p(y|x,\gamma )=\frac{\gamma ^{1/2}(x)}{(2\pi )^{3/2}}\,\exp \left[ -\frac{1}{%
2}\gamma _{ij}(x)(y^{i}-x^{i})(y^{j}-x^{j})\right] \,,  \label{p(y|x)}
\end{equation}%
and considerations of distinguishability among points (as revealed by
appropriate test particles) lead us to introduce the metric field $\gamma
_{ij}(x)$. The idea is general but was developed explicitly only for the
special case of small uncertainties, that is, for test particles that are
localized within regions much smaller than those where curvature effects
become appreciable. Situations of extreme curvature found near singularities
will not be considered here.

But there is a feature of the distinguishability distance $d\lambda $ in (%
\ref{Fisher metric}) that is very significant: it is dimensionless. Indeed,
in eq.(\ref{p(y|x)}) we can see that the metric $\gamma _{ij}(x)$ measures
spatial lengths in units of the local uncertainty: if the local uncertainty
is $\sigma (x)$, then the actual Riemannian metric is $g_{ij}(x)=\sigma
^{2}(x)\gamma _{ij}(x)$ . This immediately raises the question of how to
compare the uncertainties $\sigma (x)$ at two separate points. Information
geometry only allows one to compare the lengths of small segments at the
same place; it allows one to measure angles; it does not describe the full
geometry of space; it only describes its conformal geometry. To assign a
geometry to space we need to introduce an additional scalar field $\sigma (x)
$.

One possibility, which we pursue in the rest of this paper, is that $\gamma
_{ij}$ only describes the conformal geometry of space and that \emph{this is
all we really need}. (Entropic dynamics is defined on a space of probability
distributions, no additional structure is needed.) Perhaps the answer to the
question of how to compare uncertainties at two different locations is: Why
would we care? It is not that the irreducible uncertainty $\sigma (x)$
varies from point to point; perhaps such a comparison is objectively
meaningless and therefore unnecessary. How can we define the length of an
extended curve? Or, how can we compare distant lengths? We cannot. For most
practical purposes this does not matter because usually we are only
concerned with local distances and information geometry is quite adequate
for this restricted purpose.

But if we strongly feel that we must compare distant lengths as a tool for
reasoning, if we feel that we must define the length of curves for the sole
purpose of constructing images and pictures in order to visualize the
universe, then to satisfy this merely psychological urge, we can introduce a
field $\sigma (x)$. In this case our predictions should not depend on the
particular choice of $\sigma (x)$ which, being arbitrary, might as well be
chosen to make our models convenient and simple. The selection of $\sigma (x)
$ should be guided by purely esthetic considerations: distance should be
defined so that motion looks simple.

\section{Change}

We define the macrostate of space as a product over individual space points,%
\footnote{%
There is an assumption here that we do not need to keep track of information
about correlations among degrees of freedom at different locations.
Information about correlations may eventually turn out to be relevant
(perhaps to account for non-gravitational interactions) and could be
included in more elaborate statistical models of geometrodynamics.} 
\begin{equation}
P[\{y\}|\gamma ]=\tprod\limits_{x}p\left( y|x,\gamma \right) ~.
\label{state P}
\end{equation}%
To quantify the change from one state to another we use, once again, the
Fisher-Rao metric, but a complication arises here. The comparison between
two neighboring product states $P[\{y\}|\gamma ]$ and $P[\{y\}|\gamma
+\Delta \gamma ]$ is carried out by comparing the individual factors and we
need an explicit criterion to match factors in one state with factors in the
other. For each position $x$ in one state we must decide which is the
matching $x^{\prime }$ in the other state. We must establish a relation of
\textquotedblleft equilocality\textquotedblright . Let us provisionally
assume that a best-matching criterion has been found and that equilocal
points have been assigned the same (or \textquotedblleft
commoving\textquotedblright ) coordinates. Later we return to the question
of specifying the \textquotedblleft best-matching\textquotedblright\
criterion.

Since the state (\ref{state P}) is a product, the change from $%
P[\{y\}|\gamma ]$ to $P[\{y\}|\gamma +\Delta \gamma ]$ is a sum where the
contributions of the different degrees of freedom add in quadrature, 
\begin{equation}
\Delta L^{2}=\tsum\limits_{x}\Delta \ell ^{2}(x)~,  \label{DL2}
\end{equation}%
where $\Delta \ell ^{2}(x)$ measures the change from $p\left( y|x,\gamma
\right) $ to its equilocal counterpart $p\left( y|x,\gamma +\Delta \gamma
\right) .$ For each position $x$, we have 
\begin{equation}
\Delta \ell ^{2}(x)=G^{ij\,kl}\Delta \gamma _{ij}\Delta \gamma _{kl}\,\,,
\end{equation}%
where, using eq.(\ref{p(y|x)}), 
\begin{eqnarray}
G^{ij\,kl} &=&\int d^{3}y\,p(y|x,\gamma )\,\frac{\partial \log p(y|x,\gamma )%
}{\partial \gamma _{ij}}\,\frac{\partial \log p(y|x,\gamma )}{\partial
\gamma _{kl}}  \notag \\
&=&\frac{1}{4}\left( \gamma ^{ik}\gamma ^{jl}+\gamma ^{il}\gamma
^{jk}\right) ~.  \label{Gijkl}
\end{eqnarray}

We can write the sum in eq.(\ref{DL2}) as an integral if we note that the
density of distinguishable distributions is $\gamma ^{1/2}$. In other words,
the number of distinguishable distributions, or \textquotedblleft
distinguishable points\textquotedblright , within the coordinate interval $%
dx $ is $dx\,\gamma ^{1/2}$ ($dx$ stands for $d^{3}x$).\footnote{%
Note that since we cannot compare distant lengths or distant volumes it
makes no sense to say that $\tint_{\mathcal{R}}dx\,\gamma ^{1/2}$ measures
the volume of an extended region $\mathcal{R}$; it measures the number of
distinguishable points in $\mathcal{R}$.} Then (\ref{DL2}) is replaced by 
\begin{equation}
\Delta L^{2}=\tint dx\,\gamma ^{1/2}\Delta \ell ^{2}=\tint dx\,\gamma
^{1/2}G^{ij\,kl}\Delta \gamma _{ij}\Delta \gamma _{kl}~.  \label{DL2 a}
\end{equation}%
\emph{Two points in space count as separate only to the extent that they can
be distinguished.} The effective number of spatial degrees of freedom, that
is the number of \textquotedblleft distinguishable points\textquotedblright\
in the coordinate interval $dx$ is finite. This is neither due to an
underlying discreteness in the structure of space nor to quantum effects,
but due to the underlying intrinsic fuzziness of space.

To describe the change $\Delta \gamma _{ij}(x)$ at each location $x$ it is
convenient to introduce an arbitrary \textquotedblleft
time\textquotedblright\ parameter $t$ along the trajectory, 
\begin{equation}
\Delta \gamma _{ij}=\gamma _{ij}(t+\Delta t,x)-\gamma _{ij}(t,x)=\partial
_{t}\gamma _{ij}\Delta t~,
\end{equation}%
$\partial _{t}\gamma _{ij}$ is the \textquotedblleft
velocity\textquotedblright\ of the metric\ in the special best-matched
frame. Then eq.(\ref{DL2 a}) becomes 
\begin{equation}
\Delta L^{2}=\tint dx\,\gamma ^{1/2}G^{ij\,kl}\partial _{t}\gamma
_{ij}\partial _{t}\gamma _{kl}\Delta t^{2}~.  \label{DL2 b}
\end{equation}

Having computed the change in the special commoving frame where equilocal
points have the same coordinates we now switch to an arbitrary coordinate
frame where equilocal points at $t$ and $t+\Delta t$ have coordinates $x^{i}$
and $\tilde{x}^{i}=x^{i}-\beta ^{i}(x)\Delta t$ respectively; equilocal
points are \textquotedblleft shifted\textquotedblright\ by $\beta ^{i}\Delta
t$. Under the infinitesimal shift $\tilde{x}^{i}=x^{i}-\beta ^{i}(x)\Delta t$
the metric at $t+\Delta t$ transforms into $\tilde{\gamma}_{ij},$ 
\begin{equation}
\gamma _{ij}(t+\Delta t,x)=\tilde{\gamma}_{ij}(t+\Delta t,x)-\left( \nabla
_{i}\beta _{j}+\nabla _{j}\beta _{i}\right) \Delta t~,
\end{equation}%
where $\nabla _{i}\beta _{j}=\partial _{i}\beta _{j}-\Gamma _{ij}^{k}\beta
_{k}$ is the covariant derivative associated to the metric $\gamma _{ij}$.
In the new frame, setting $\tilde{\gamma}_{ij}(t+\Delta t,x)-\gamma
_{ij}(t,x)=\Delta \gamma _{ij}$, the change in $\gamma _{ij}$ between
equilocal points is expressed as

\begin{equation}
\Delta _{\beta }\gamma _{ij}=\Delta \gamma _{ij}-\left( \nabla _{i}\beta
_{j}+\nabla _{j}\beta _{i}\right) \Delta t~,  \label{Dbeta gamma}
\end{equation}%
or, $\Delta _{\beta }\gamma _{ij}=\dot{\gamma}_{ij}\Delta t$, where 
\begin{equation}
\dot{\gamma}_{ij}\overset{\limfunc{def}}{=}\partial _{t}\gamma _{ij}-\nabla
_{i}\beta _{j}-\nabla _{j}\beta _{i}~.  \label{gamma dot}
\end{equation}%
In terms of the transformed coordinates the change $\Delta L^{2}$ retains
the same form as before, eq.(\ref{DL2 b}), except that the new best-matched
velocities $\dot{\gamma}_{ij}$ are the coordinate velocities $\partial
_{t}\gamma _{ij}$ suitably \textquotedblleft corrected\textquotedblright\ by
the shift $\beta ^{i}$, 
\begin{equation}
\Delta _{\beta }L^{2}=\tint dx\,\gamma ^{1/2}G^{ij\,kl}\dot{\gamma}_{ij}\dot{%
\gamma}_{kl}\Delta t^{2}~.  \label{DL2 c}
\end{equation}%
Note that $\Delta L^{2}$ depends only on the initial and final states and is
invariant under the reparametrization of time $t\rightarrow t^{\prime
}=f(t,x)$.

Now we address the problem of specifying the best-matching criterion. For
given velocities $\partial _{t}\gamma _{ij}$ our estimate $\Delta _{\beta
}L^{2}$ of the actual change $\Delta L^{2}$ can be artificially altered by
different choices of the shift $\beta ^{i}$. We have to decide which values
of $\beta ^{i}$ provide the best equilocality match.

The problem of selecting the optimal shift can be tackled as a problem of
inference: the \textquotedblleft prior\textquotedblright\ state of
information is described by the earlier distribution $P_{t}=P[\{y\}|\gamma ]$%
, and we are given the new information that the \textquotedblleft
posterior\textquotedblright\ state belongs to the later \textquotedblleft
trial\textquotedblright\ family of distributions $P_{t+\Delta
t}=P[\{y\}|\gamma +\Delta \gamma ]$. The trial distributions are essentially
identical except for diffeomorphisms -- the spatial shifts $\beta ^{i}\Delta
t$. Which one do we choose? \emph{We choose the distribution that does the
least violence to our prior beliefs while fully accommodating the new
information.} Phrased in this way it is clear that this is the kind of
question the method of maximum entropy was designed to answer: \emph{Best
matching reflects the least change.}

The actual change $\Delta L^{2}$ between the two successive states is
obtained [using the property in eq.(\ref{Sdlambda})] either by maximizing
the appropriate relative entropy $S[P_{t+\Delta t}|P_{t}]$ or by minimizing
the corresponding $\Delta _{\beta }L^{2}$, $S[P_{t+\Delta t}|P_{t}]=-\Delta
_{\beta }L^{2}/2$, over all choices of $\beta ^{i}$, 
\begin{equation}
\Delta L^{2}={}\underset{\beta }{\min }\;\Delta _{\beta }L^{2}\,.
\end{equation}%
Vary with respect to $\beta $, 
\begin{equation}
\delta \left( \Delta _{\beta }L^{2}\right) =2\int dx\,\gamma ^{1/2}G^{ij\,kl}%
\dot{\gamma}_{ij}\delta \dot{\gamma}_{kl}~\Delta t^{2}=0~.
\end{equation}%
Next use $\delta \dot{\gamma}_{kl}=-\nabla _{k}\delta \beta _{l}-\nabla
_{l}\delta \beta _{k}$ and integrate by parts to get%
\begin{equation}
\nabla _{l}\left( 2G^{ij\,kl}\dot{\gamma}_{ij}\right) =0~\quad \text{or}%
\quad \nabla _{l}\dot{\gamma}^{kl}=0~,  \label{best matching}
\end{equation}%
where we used eq.(\ref{Gijkl}) and 
\begin{equation}
\dot{\gamma}^{kl}=\partial _{t}\gamma ^{kl}+\nabla ^{k}\beta ^{l}+\nabla
^{l}\beta ^{k}~.  \label{gamma dot up}
\end{equation}

Eqs.(\ref{best matching}) are the differential equations that determine the
shift $\beta ^{i}$ that establishes the best matching and equilocality
between the given initial and final geometries $\gamma _{ij}$ and $\gamma
_{ij}+\Delta \gamma _{ij}$. Alternatively, we can consider these equations
as constraints on the allowed change $\Delta \gamma _{ij}=\partial
_{t}\gamma _{ij}\Delta t$ for a given shift $\beta ^{i}$.

\section{Entropic dynamics}

The dynamical question is \textquotedblleft Given initial and final states,
what trajectory is the system expected to follow?\textquotedblright\ The
answer \cite{Caticha03b}\cite{Caticha02} follows from the implicit
assumption that there exists a continuous trajectory. This reduces the
problem of studying large changes to the simpler problem of studying small
changes. 

Consider the short segment of the trajectory between the states $P_{t}$ and $%
P_{t+\Delta t}$. The idea is that in going from one to the other the system
must pass through a halfway point, and also through a state that lies a
third of the way, and so on. More generally, the trajectory is composed of
states such that having travelled a distance $dL$ from the initial $P_{t}$,
there remains a distance $\omega dL$ to the final $P_{t+\Delta t}$, with $%
0<\omega <\infty $. The trajectory is the set of states obtained as $\omega $
sweeps from $0$ to $\infty $. 

However, in the case of geometrodynamics we know much more than just that
the product state eq.(\ref{state P}) must evolve through a continuous
sequence of intermediate states. \emph{We also know that each and every one
of the individual factors must evolve continuously through a sequence of
intermediate states to reach the corresponding final state.} This means that
instead of one parameter $\omega $ there are many such parameters, one for
each position $x$. In other words, the intermediate states $P_{{}\omega }$
interpolating between the initial $P_{t}$ and the final $P_{t+\Delta t}$
should be labeled by a function $\omega (x)=w\zeta (x)$ where $\zeta (x)$ is
a fixed positive function and the parameter $w$ varies from $0$ to $\infty $.

There is no single trajectory; each choice of the function $\zeta (x)$
defines one possible trajectory. In a sense, the system follows many
alternative paths simultaneously -- this is Wheeler's many-fingered time --
and physical predictions are independent of the choice of the arbitrary
function $\zeta (x)$. The path-independence is very significant because the
product state $P_{t}$ provides us with the only definition of what an
\textquotedblleft instant\textquotedblright\ is, of what state $p\left(
y|x^{\prime }\right) $ of a distant test particle at $x^{\prime }$ we can
agree to call simultaneous with a certain state $p\left( y|x\right) $ of the
test particle at $x$. Therefore, if there is no unique sequence of
intermediate states, then there is no unique, absolute definition of
simultaneity. We see here a \textquotedblleft foliation\textquotedblright\
invariance, a rudimentary form of local Lorentz invariance.

Let $t$ be the \textquotedblleft time\textquotedblright\ parameter labeling
successive intermediate states. The initial state is $\gamma
_{ij}(t,x)=\gamma _{ij}(x)$, the final state is $\gamma _{ij}(t+\Delta
t,x)=\gamma _{ij}(x)+\Delta \gamma _{ij}(x)$, and the intermediate states
are of the form $\gamma _{ij}(t+dt,x)=\gamma _{ij}(x)+d\gamma _{ij}(x)$. For
appropriate choices of the shift the best-matched changes corresponding to $%
\Delta \gamma _{ij}$ and $d\gamma _{ij}$ are given by eq.(\ref{Dbeta gamma})
and%
\begin{equation}
d_{\beta }\gamma _{ij}=d\gamma _{ij}-\left( \nabla _{i}\beta _{j}+\nabla
_{j}\beta _{i}\right) dt~.  \label{dbeta gamma}
\end{equation}

To determine the intermediate state $P_{t+dt}$ one varies $d\gamma _{ij}$ to
maximize the relative entropy 
\begin{equation}
S[P_{t+dt}|P_{t}]=-\frac{1}{2}dL^{2}=-\frac{1}{2}\int dx\,\gamma ^{1/2}d\ell
^{2}(x)~,
\end{equation}%
subject to independent constraints at each point $x$. For each of the
factors in the product state $P_{t+dt}$ we require that if the distance to
the initial state is $d\ell (x)$ then the distance that remains to be
covered to reach the final state is $d\ell _{f}(x)=\omega (x)d\ell (x)$
where 
\begin{equation}
d\ell ^{2}(x)=G^{ij\,kl}d_{\beta }\gamma _{ij}d_{\beta }\gamma _{kl}~,
\end{equation}%
and 
\begin{equation}
d\ell _{f}^{2}(x)=G^{ij\,kl}\left[ \Delta _{\beta }\gamma _{ij}-d_{\beta
}\gamma _{ij}\right] \left[ \Delta _{\beta }\gamma _{kl}-d_{\beta }\gamma
_{kl}\right] ~.
\end{equation}%
Introducing Lagrange multipliers $\lambda (x)$, the basic variational
principle of entropic dynamics is 
\begin{equation}
0=\delta \int dx\,\gamma ^{1/2}\left[ d\ell ^{2}+\lambda \left( d\ell
_{f}^{2}-\omega ^{2}d\ell ^{2}\right) \right] ~.  \label{ent dyn}
\end{equation}%
Variations of $d\gamma _{kl}$ give 
\begin{equation}
d_{\beta }\gamma _{ij}(x)=\chi (x)\Delta _{\beta }\gamma _{ij}(x)\quad \text{%
where}\quad \chi (x)=\frac{\lambda (x)}{1+\lambda (x)\left( 1-\omega
^{2}(x)\right) }\,.
\end{equation}%
The Lagrange multipliers $\lambda (x)$ are determined so that the
constraints $d\ell _{f}=\omega d\ell $ hold. We get 
\begin{equation}
d\ell (x)=\chi \Delta \ell (x)\quad \text{and}\quad \chi (x)=\frac{1}{%
1+\omega (x)}~\,,
\end{equation}%
and conclude that the selected intermediate state $d\gamma _{ij}$ is such
that 
\begin{equation}
d\ell (x)+d\ell _{f}(x)=\Delta \ell (x)~,  \label{ED}
\end{equation}%
which means that the metric at the point $x$ (the metric $\gamma $) evolves
along geodesics in its individual configuration space. Degrees of freedom at
different locations do not, however, evolve independently of each other;
they are coupled through the diffeomorphism constraint, eq.(\ref{best
matching}), which decides, at each moment in time, which spatial points are
equilocal. Note that the trajectory described by (\ref{ED}) is explicitly
independent of $\omega (x)$; this is foliation invariance.

Having derived a model of statistical geometrodynamics by applying standard
rules of inference to the information codified in the states of the system,
we can now summarize the dynamics by introducing an action that leads to the
same equations of motion. The proposed action is 
\begin{equation}
J=\int_{t_{i}}^{t_{f}}dt\,\int dx\,\gamma ^{1/2}\left( G^{ij\,kl}\dot{\gamma}%
_{ij}\dot{\gamma}_{kl}\right) ^{1/2}.  \label{action J}
\end{equation}%
Our next step should be to explore the consequences of this statistical
geometrodynamics and establish the relation, if any, between this theory and
Einstein's General Relativity, but this is a subject for future work.

\section{Conclusions and some comments}

The model of statistical geometrodynamics (SGD) developed here combines two
basic ideas. First, the geometry of space is of statistical origin and is
explained in terms of the distinguishability metric of Fisher and Rao.
Second, the dynamics of this geometry is derived purely on the basis of
principles of inference; there is no need to postulate additional
\textquotedblleft laws of nature.\textquotedblright\ 

The similarities with the general theory of relativity (GR) suggest that GR
can be obtained in some appropriate limit. For example, just as in GR the
dynamical degrees of freedom are those that specify the conformal geometry
of space \cite{York71}. The best-matching condition corresponds to the
diffeomorphism constraint in the Hamiltonian formulation of GR \cite%
{Arnowitt62}. There is no reference to an external time; there is a natural
intrinsic time defined by the change of the system itself which, just as in
GR, can only be obtained after the equations of motion are solved \cite%
{Baierlein62}. Despite being derived by maximizing entropies the theory is
time reversible.

Perhaps the feature of SGD that does most violence to our intuition is its
scale invariance. The scale factor $\sigma (x)$ needed to assign a
Riemannian geometry to space is arbitrary and its choice should be dictated
by convenience. This gauge invariance can be used to great advantage. The
essence of the dynamics of GR lies in the embeddability of space in
spacetime: any model that uses only the metric tensor to describe the
changing geometry of space as it evolves in spacetime is equivalent to GR 
\cite{Hojman76}. I conjecture that the $\sigma (x)$ can be chosen so that
the evolving geometry of space sweeps out a four-dimensional spacetime --
which amounts to choosing the gauge so that the appropriate Gauss-Codazzi
equations are satisfied. In this particular gauge SGD should coincide with
GR, in other words, we will have accomplished our goal of deriving
macroscopic GR from a more basic microscopic statistical theory.

If true, statistical geometrodynamics would have a number of implications
for physics. Perhaps the most interesting are the revision it requires of
the notion of distance, the statistical structure of both time and space,
and the recognition that spacetime is not a fundamental notion. The
statistical nature of geometry could provide mechanisms that would eliminate
the infinities pervading quantum field theories, either through decoherence
or through the finite number of distinguishable points within a finite
volume. Furthermore, it would make the Lorentz and CPT symmetries have only
statistical validity and it might bear on the subject of CP violation and
matter-antimatter asymmetry. On the other hand, the scale invariance might
be relevant to cosmological issues such as the early inflation and the late
accelerated expansion of the universe.


\begin{thebibliography}{99}
\bibitem{Cox46} R. T. Cox: Am. J. Phys. \textbf{14}, 1 (1946); \emph{The
Algebra of Probable Inference} (Johns Hopkins, Baltimore, 1961).

\bibitem{Jaynes03} E. T. Jaynes, \emph{Probability Theory: The Logic of
Science} (Cambridge U. Press, Cambridge, 2003).

\bibitem{Jaynes57} E. T. Jaynes: Phys. Rev. \textbf{106}, 620 and \textbf{108%
}, 171 (1957); \emph{E. T. Jaynes: Papers on Probability, Statistics and
Statistical Physics}, ed. by R. D. Rosenkrantz (Reidel, Dordrecht, 1983).

\bibitem{ShoreJohnson80} J. E. Shore and R. W. Johnson: IEEE Trans. Inf.
Theory \textbf{IT-26}, 26 (1980).

\bibitem{Skilling88} J. Skilling: `The Axioms of Maximum Entropy'. In: \emph{%
Maximum-Entropy and Bayesian Methods in Science and Engineering}, ed. by G.
J. Erickson and C. R. Smith (Kluwer, Dordrecht, 1988).

\bibitem{Caticha03a} A. Caticha, \textquotedblleft Relative Entropy and
Inductive Inference,\textquotedblright\ in \emph{Bayesian Inference and
Maximum Entropy Methods in Science and Engineering}, ed. by G. Erickson and
Y. Zhai, AIP Conf. Proc. \textbf{707}, 75 (2004)
(arXiv.org/abs/physics/0311093).

\bibitem{Caticha98} A. Caticha: Phys. Lett. \textbf{A244}, 13 (1998); Phys.
Rev. \textbf{A57}, 1572 (1998); Found. Phys. \textbf{30}, 227 (2000)
(arXiv.org/abs/quant-ph/9810074).

\bibitem{Amari00} S. Amari and H. Nagaoka, \emph{Methods of Information
Geometry} (Am. Math. Soc./Oxford U. Press, Providence, 2000).

\bibitem{Rodriguez02} C. C. Rodr\'{\i}guez, \textquotedblleft Entropic
Priors for Discrete Probabilistic Networks and for Mixtures of Gaussian
Models\textquotedblright\ in \emph{Bayesian Inference and Maximum Entropy
Methods in Science and Engineering}, ed. by R. L. Fry, AIP Conf. Proc. 
\textbf{617}, 410 (2002); \textquotedblleft The ABC of Model Selection: AIC,
BIC, and the New CIC\textquotedblright , in these Proceedings and also at
http://omega.albany.edu:8008/.

\bibitem{Caticha03b} A. Caticha, \textquotedblleft Towards a Statistical
Geometrodynamics\textquotedblright\ in \emph{Decoherence and Entropy in
Complex Systems} ed. by H.-T. Elze (Springer Verlag, 2004)
(arXiv.org/abs/gr-qc/0301061).

\bibitem{Fisher25} R. A. Fisher: Proc. Cambridge Philos. Soc. \textbf{122},
700 (1925); C. R. Rao: Bull. Calcutta Math. Soc. \textbf{37}, 81 (1945).

\bibitem{Cencov81} N. N. \v{C}encov: \emph{Statistical Decision Rules and
Optimal Inference}, Transl. Math. Monographs, vol. 53, Am. Math. Soc.
(Providence, 1981); L. L. Campbell: Proc. Am. Math. Soc. \textbf{98}, 135
(1986).

\bibitem{York71} J. W. York, Phys. Rev. Lett. \textbf{26}, 1656 (1971);
Phys. Rev. Lett. \textbf{28}, 1082 (1972); Phys. Rev. Lett. \textbf{82},
1350 (1999).

\bibitem{Caticha01} A. Caticha, \textquotedblleft Change, Time and
Information Geometry\textquotedblright\ in \emph{Bayesian Methods and
Maximum Entropy in Science and Engineering}, ed. by A. Mohammad-Djafari, AIP
Conf. Proc. \textbf{568}, 72 (2001) (arXiv.org/abs/math-ph/0008018)

\bibitem{Barbour94} J. Barbour: Class. Quant. Grav. \textbf{11}, 2853 (1994).

\bibitem{Arnowitt62} R. Arnowitt, S. Deser and C. W. Missner,
\textquotedblleft The Dynamics of General Relativity\textquotedblright\ in 
\emph{Gravitation: an Introduction to Current Research}, ed. by L. Witten
(Wiley, New York, 1962) (arXiv.org/abs/gr-qc/0405109).

\bibitem{Caticha02} A. Caticha, \textquotedblleft Entropic
Dynamics\textquotedblright\ in \emph{Bayesian Inference and Maximum Entropy
Methods in Science and Engineering}, ed. by R. L. Fry, AIP Conf. Proc. 
\textbf{617}, 302 (2002)  (arXiv.org/abs/gr-qc/0109068).

\bibitem{Baierlein62} R. F. Baierlein, D. H. Sharp and J. A. Wheeler: Phys.\
Rev. \textbf{126}, 1864 (1962).

\bibitem{Hojman76} S. A. Hojman, K. Kuchar and C. Teitelboim: Ann. Phys. 
\textbf{96}, 88 (1976); K. Kuchar, J. Math. Phys. \textbf{15}, 708 (1974).
\end{thebibliography}
\end{document}